\documentclass[letterpaper,twocolumn,english,floatfix, aps, prl]{revtex4}
\usepackage[T1]{fontenc}
\usepackage[latin1]{inputenc}
\usepackage{babel}
\usepackage{graphics}

\makeatletter

\providecommand{\LyX}{L\kern-.1667em\lower.25em\hbox{Y}\kern-.125emX\@}
\let\SF@@footnote\footnote
\def\footnote{\ifx\protect\@typeset@protect
    \expandafter\SF@@footnote
  \else
    \expandafter\SF@gobble@opt
  \fi
}
\expandafter\def\csname SF@gobble@opt \endcsname{\@ifnextchar[
  \SF@gobble@twobracket
  \@gobble
}
\edef\SF@gobble@opt{\noexpand\protect
  \expandafter\noexpand\csname SF@gobble@opt \endcsname}
\def\SF@gobble@twobracket[#1]#2{}

\usepackage{graphicx}
\usepackage{dcolumn}
\usepackage{bm}

\makeatother
\begin{document}

\title{Random-field Ising model on complete graphs and trees }

\author{R. Dobrin}

\email{dobrin@pa.msu.edu}

\author{J.H. Meinke}

\author{P.M. Duxbury}

\affiliation{Dept. of Physics \& Astronomy and Center for Fundamental Materials
Research,\\
 Michigan State University, East Lansing, MI 48824, USA}

\begin{abstract}
We present exact results for the critical behavior of the RFIM on
complete graphs and trees, both at equilibrium and away from equilibrium,
i.e., models for hysteresis and Barkhausen noise. We show that for stretched
exponential and power law distributions of random fields the behavior
on complete graphs is non-universal, while the behavior on Cayley trees is universal 
even in the limit of large co-ordination. 
\end{abstract}
\maketitle
The central issue in the equilibrium random field Ising model(RFIM)
is the nature of the phase transition from the ferromagnetic state
at weak disorder to the frozen paramagnetic state at high disorder.
The existence and universality class of the RFIM transition, is key
as the best experimental tests of RFIM theory are diluted antiferromagnets
in a field, which are believed to be in the same universality class
as the RFIM \cite{fishman79a}. After some controversy it was rigorously
demonstrated that the RFIM transition occurs at a finite width of
the distribution in three dimensions \cite{imbrie84a} and at an infinitesimal
width in one and two dimensions. Moreover, Aharony \cite{aharony78a}
showed that within mean field theory at low temperatures, the transition
is first order for bimodal disorder distributions but second order
for unimodal distributions. Numerical studies at zero temperature
suggest that in four dimensions the bimodal case is first order and
the Gaussian case is second order. The analysis in three dimensions
is less conclusive \cite{swift97a}. The difference between the Gaussian
and bimodal cases has been attributed to percolative effects \cite{swift94a}.
We have recently shown that at zero temperature, the mean-field theory
is \textit{non-universal} \cite{duxbury01a} in the sense that the
order parameter exponent may vary continuously with the disorder.
Exact optimization calculations \cite{rieger97,alava01a} in three dimensions
have also suggested that the correlation
length exponent, as deduced from finite size scaling, is non-universal
\cite{angles97a}. 

Motivated by the fact that the RFIM is non-universal
within mean-field theory for the stretched exponential
distribution, we have analyzed the the RFIM on complete
graphs with disorder distribution, \( (\delta h/|h|)^{x} \) (\( 0<x<1 \),
\( |h|<\delta h \)).  We find that this distribution is 
anomalous in the sense that this sort of disorder never
destroys the spontaneously magnetized state, at least within
mean-field theory. The behavior of the RFIM on complete graphs is thus
quite varied and anomalous.  To determine whether this
non-universality extends to other lattices, 
we have analyzed the zero temperature RFIM
on a \textit{Bethe} lattice for the stretched exponential and
power law distributions of disorder.
We prove that \textit{the Bethe lattice is universal}, 
 provided the transition is second order, 
even in the limit of large co-ordination. This
is surprising since in this limit the Bethe lattice usually approaches
the mean-field limit. 

We also extend the results outlined above to the non-equilibrium case.
Ground state calculations of hysteresis and Barkhausen noise in the
RFIM have demonstrated that the spin avalanches are controlled by
the equilibrium RFIM critical point \cite{dahmen96a,sethna01a}. It
is thus not surprising, and we confirm,
 that the magnetization jump in the hysteresis
loop is non-universal for the stretched exponential disorder
distribution. The integrated avalanche distribution also has a
non-universal exponent due to the non-universality of the order parameter.
But the {}``differential\char`\"{} mean-field avalanche exponent
is universal even in cases where the order parameter exponent is not.
In contrast, as expected from the equilibrium results, 
the Bethe lattice exhibits universal non-equilibrium critical
behavior.

The Hamiltonian of the random-field Ising model is, \begin{equation}
{\mathcal{H}}=-\sum _{ij}J_{ij}S_{i}S_{j}-\sum _{i}(H+h_{i})S_{i},
\end{equation}
 where the exchange is ferromagnetic (\( J_{ij}>0 \)) and the fields
\( h_{i} \) are random and uncorrelated. In the non-equilibrium problem
we sweep the applied uniform field, \( H \), from \( -\infty  \)
to \( \infty  \) and monitor the magnetization at a fixed \( J_{ij}=J \)
and for a fixed disorder configuration \( \{h_{i}\} \). This model
has been proposed as a model for Barkhausen noise by Dahmen et al.~\cite{dahmen96a}.
The local effective field responsible for a spin-flip is \begin{equation}
h_{i}^{eff}=J\sum _{j\not =i}S_{j}+h_{i}+H
\end{equation}
 The condition for a spin to flip is that \( h_{i}^{eff}>0 \). The
random fields are drawn from a specified distribution \( \rho (h) \).
To test universality, we use the following distributions which are
defined on the interval -\( \delta h\leq h\leq \delta h \), \begin{equation}
\label{eq:poly_dist_1}
\rho _{1}(h)=\frac{y+1}{2y\, \delta h}\left[ 1-\left( \frac{|h|}{\delta h}\right) ^{y}\right] \quad \, \, 0<y<\infty 
\end{equation}
 and \begin{equation}
\label{eq:poly_dist_2}
\rho _{2}(h)=\frac{y+1}{2\, \delta h}\left( \frac{|h|}{\delta h}\right) ^{y}\quad -1<y<\infty 
\end{equation}
 We have shown that \( \rho _{1} \), which is the low field expansion
of a stretched exponential disorder distribution, leads to non-universality
in the ground state of the equilibrium mean-field RFIM \cite{duxbury01a}.
Here we extend that result to the non-equilibrium case. We then show
that the distribution \( \rho _{2} \) destroys the RFIM phase transition,
in mean-field theory and on trees, for \( -1<y<0 \).

First we discuss the behavior of the ground state of the zero-temperature,
mean-field RFIM. The magnetization is given by \begin{equation}
m=-\int _{-\infty }^{h_{c}(m)}\rho (h)dh+\int _{h_{c}(m)}^{\infty }\rho (h)dh
\end{equation}
 were \( h_{c}(m)=-Jm-H \). The energy at a given magnetization is
\begin{equation}
\label{eq:energy}
E(m)={Jm^{2}\over 2}-\int _{-\infty }^{\infty }|h|\rho (h)dh+2\int _{0}^{h_{c}}h\rho (h)dh.
\end{equation}

Extremizing with respect to the order parameter, \( m \), yields
the ground-state mean-field equation, \begin{equation}
\label{eq:ground_state_mfe}
m_{e}=2\int _{0}^{Jm_{e}+H}\rho (h)dh
\end{equation}

The non-equilibrium critical points are found from the susceptibility
\( \chi =\partial m/\partial H \), which from (\ref{eq:ground_state_mfe})
is given by, \begin{equation}
\label{eq:susceptibility}
\chi ={2 \rho (Jm+H)\over 1-2J \rho (Jm+H)}
\end{equation}
 The avalanche distribution, \( d(s,t) \) that gives the probability
of finding an avalanche of size \( s \) at parameter value \( t \),
is found using a Poisson statistics argument \cite{dahmen96a}, which
yields, \begin{equation}
\label{eq:avalanche_distribution}
d(s,t)\sim s^{-\tau }e^{-t^{2}s}=s^{-\tau }g(s^{\sigma }t),
\end{equation}
 where \( g(x) \) is a scaling function and \( t=1-2 J \rho (Jm_{e}+H) \).
Experimentally, it is more natural to make a histogram of all avalanches
up to the critical applied field at which the magnetization changes
sign. This {}``integrated\char`\"{} distribution behaves as, \begin{equation}
D(s,\delta h)=s^{-\tau -\sigma \beta \delta }g(s^{\sigma }r)
\end{equation}
 where \( r=|\delta h-\delta h_{c}| \). For a Gaussian distribution
of disorder, \( \beta =1/2 \), \( \sigma =1/2 \), \( \tau =3/2 \).
We have shown, however, that in the ground state for the distribution
(\ref{eq:poly_dist_1}), the equilibrium order parameter exponent,
\( \beta =1/y \). In contrast it is evident from Eq.~(\ref{eq:avalanche_distribution})
that the exponents \( \sigma  \) and \( \tau  \) are universal.
The non-universality in non-equilibrium behavior arises
in the magnetization jump and in the shape of the
non-equilibrium phase boundary, as we now demonstrate.
 Consider the distribution (\ref{eq:poly_dist_1}).
Integrating (\ref{eq:ground_state_mfe}) yields the mean field equations,
\begin{equation}
\label{eq:mfe_of_distro1_greater}
m={\frac{y+1}{y}}(\overline{J}m+\overline{H})-{\frac{1}{y}}|\overline{J}m+\overline{H}|^{y+1}
\end{equation}
 for \( Jm+H>0 \), and \begin{equation}
\label{eq:mfe_of_distro1_less}
m={\frac{y+1}{y}}(\overline{J}m+\overline{H})+{\frac{1}{y}}|\overline{J}m+\overline{H}|^{y+1}
\end{equation}
 for \( Jm+H<0 \). Here we have defined, \( \overline{J}=J/\delta h \),
\( \overline{H}=H/\delta h \). Setting \( \overline{H}=0 \) in either
(\ref{eq:mfe_of_distro1_greater}) or (\ref{eq:mfe_of_distro1_less})
yields the equilibrium magnetization \cite{duxbury01a}, \begin{equation}
m_{eq}={1\over \overline{J}}\left[ {y+1}\right] ^{1/y}\left[ 1-{y\over \overline{J}(y+1)}\right] ^{1/y}
\end{equation}
 At the critical point, the magnetization scales with the magnetic
field as \( m_{e}(r=0,H)\sim H^{1/\delta } \). From Eq.~(\ref{eq:mfe_of_distro1_less})
it is evident that \( \delta =y+1 \).
 The susceptibility \( \chi =\partial m/\partial \overline{H} \)
diverges when the barrier between the two local magnetization minima
of the ground state energy ceases to exist. From (\ref{eq:susceptibility}),
we have \begin{equation}
\chi =\frac{(y+1)[1-(\overline{J}m+\overline{H})^{y}]}{y-(y+1)\overline{J}[1-(\overline{J}m+\overline{H})^{y}]}.
\end{equation}
 and the critical condition \begin{equation}
y=(y+1)\overline{J}[1-(\overline{J}m_{neq}+\overline{H_{c}})^{y}].
\end{equation}
 This equation has the simple solution, \begin{equation}
\label{eq:x_c}
x_{c}=\overline{J}m_{neq}+\overline{H_{c}}=\left[ 1-\frac{y}{\overline{J}(y+1)}\right] ^{1/y}.
\end{equation}
 Substituting (\ref{eq:x_c}) into (\ref{eq:mfe_of_distro1_greater}),
we find that the non-equilibrium magnetization jump is positive and
has the value \begin{equation}
m_{neq}=\left[ 1+\frac{1}{\overline{J}(y+1)}\right] \left[ 1-\frac{y}{\overline{J}(y+1)}\right] ^{1/y},
\end{equation}
 for \( H\rightarrow H_{c}^{+} \). Substituting this into (\ref{eq:x_c}),
the critical field is found to be, \begin{equation}
H_{c}=-J\left[ 1-\frac{y\, \, \delta h}{J(y+1)}\right] ^{1+1/y}.
\end{equation}

\begin{figure}
{\centering
\resizebox*{0.68\columnwidth}{!}{\rotatebox{270}{\includegraphics{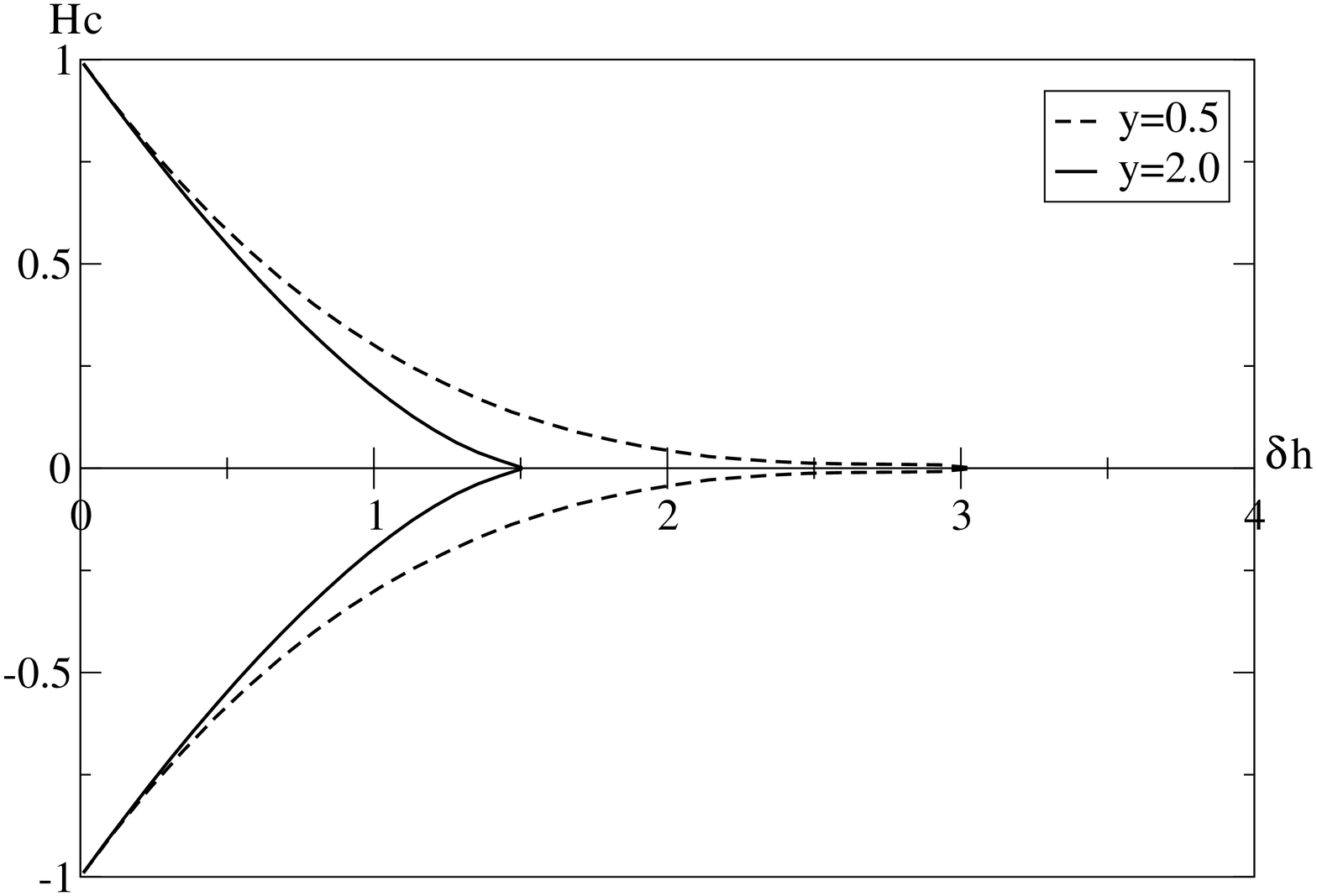}}} \par}

{\centering
\resizebox*{0.68\columnwidth}{!}{\rotatebox{270}{\includegraphics{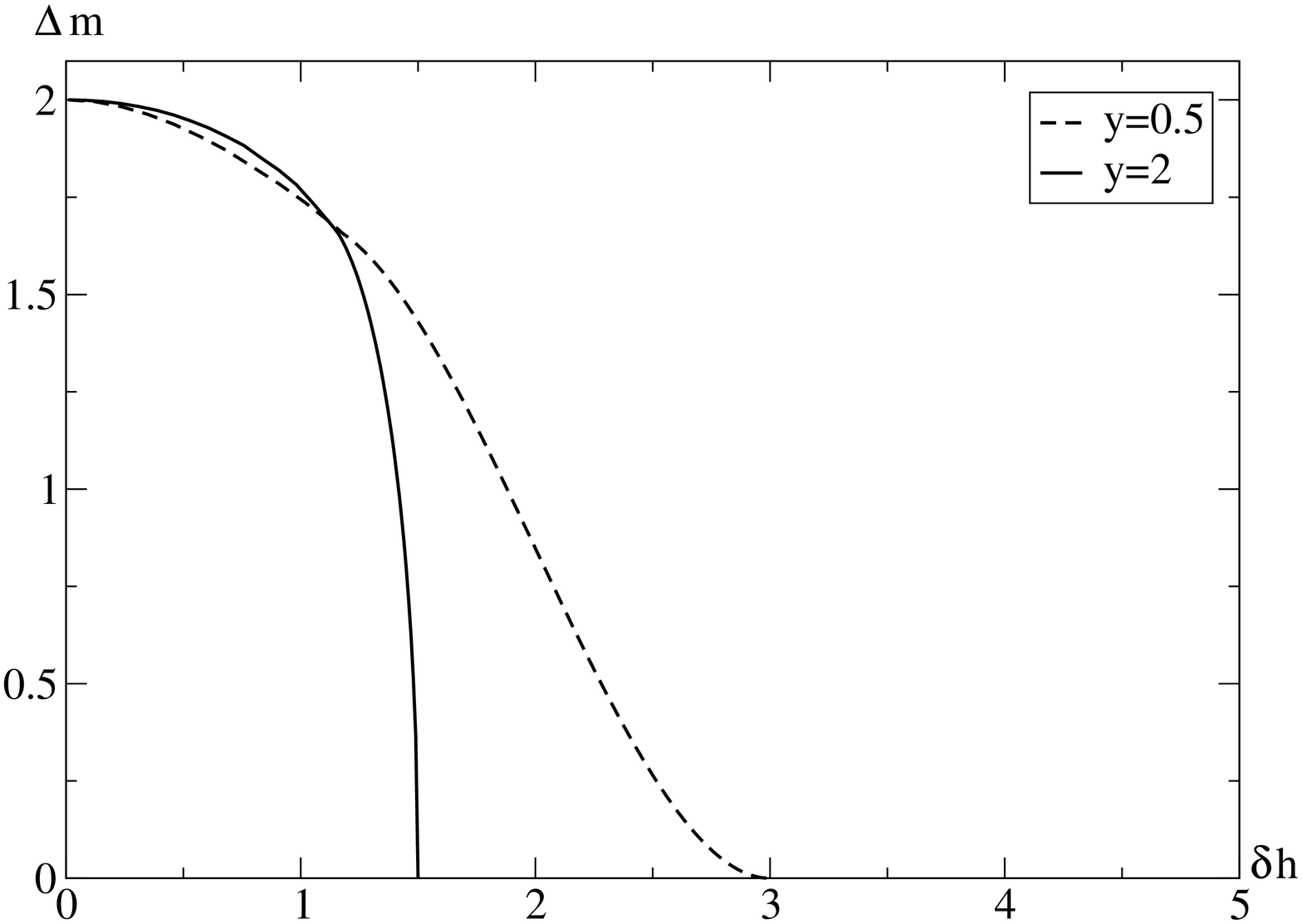}}} \par}

\caption{\textbf{Top:} The phase diagram of the non-equilibrium RFIM MFT using
the disorder distribution (\ref{eq:poly_dist_1}), and \textbf{Bottom:}
The magnetization jump at the phase boundary. In these figures, we
took the exchange constant \protect\( J=1\protect \). The dotted
line is for \protect\( y=0.5\protect \) while the solid line is for
\protect\( y=2\protect \). Note that at the equilibrium critical
disorder, \protect\( \delta h_{c}\protect \), the hysteresis loop
disappears. }

\label{fig1}
\end{figure}
 This negative critical field is expected when starting with the positive
magnetized state. By symmetry, the negative magnetization solution
is at \( -H_{c} \). The value of the magnetization at that point
is \( -m_{neq} \). Note that \( |m_{neq}| \) is \textit{not} the
size of the magnetization jump in the hysteresis loop. The jump in
magnetization in the hysteresis loop is \( \delta m_{hyst}=|m_{neq}|+m(|H_{c}|) \),
where \( m(|H_{c}|) \) is found by solving Eq.~(\ref{eq:mfe_of_distro1_less}).
The critical exponent associated with the jump in magnetization is
determined by the behavior of the distribution \( \rho (h) \) at
small fields, so that the critical exponents found here apply to distributions
of the form \( \rho (h)=exp(-(|h|/H)^{y}) \). For \( y<1 \) these
are the stretched exponential distributions ubiquitous in glasses,
while for \( y>2 \) they are more concentrated near the origin.

Now we briefly consider the distribution \( \rho _{2}(h) \) given
in Eq.~(\ref{eq:poly_dist_2}). For \( y>0 \) this distribution
is bimodal and it is easy to confirm the conclusion of Aharony \cite{aharony78a}
that the transition is first order. However the cases \( -1<y<0 \)
are more interesting. In these cases the disorder is dominated by
small random fields, as the distribution is singular at the origin.
It is easy to carry out the mean-field calculation (\ref{eq:ground_state_mfe})
with the result, \begin{equation}
\label{eq:magnetization_dist2}
m_{eq}=\left( \frac{\delta h}{J}\right) ^{1+1/y}\quad \delta h>J
\end{equation}
 By comparing the energies of \( E(m=0) \), \( E(m=1) \) and \( E(m_{eq}) \)
(using Eq.~(\ref{eq:energy})), we find that for \( \delta h<J \),
the ground state is fully magnetized, while for \( \delta h>J \)
the ground state has magnetization (\ref{eq:magnetization_dist2}).
The interesting feature of the result (\ref{eq:magnetization_dist2})
is that \textit{there is no phase transition} at finite \( \delta h \),
and the system is always ordered. The disorder distribution (\ref{eq:poly_dist_2})
thus destroys the ground state phase transition, due to the large
number of small random fields.

Now we determine whether the non-universal results found above for
the mean-field theory extend to the ground state of the RFIM on a
Cayley tree. 
\begin{figure}
{\centering \resizebox*{0.68\columnwidth}{!}{
\rotatebox{270}{\includegraphics{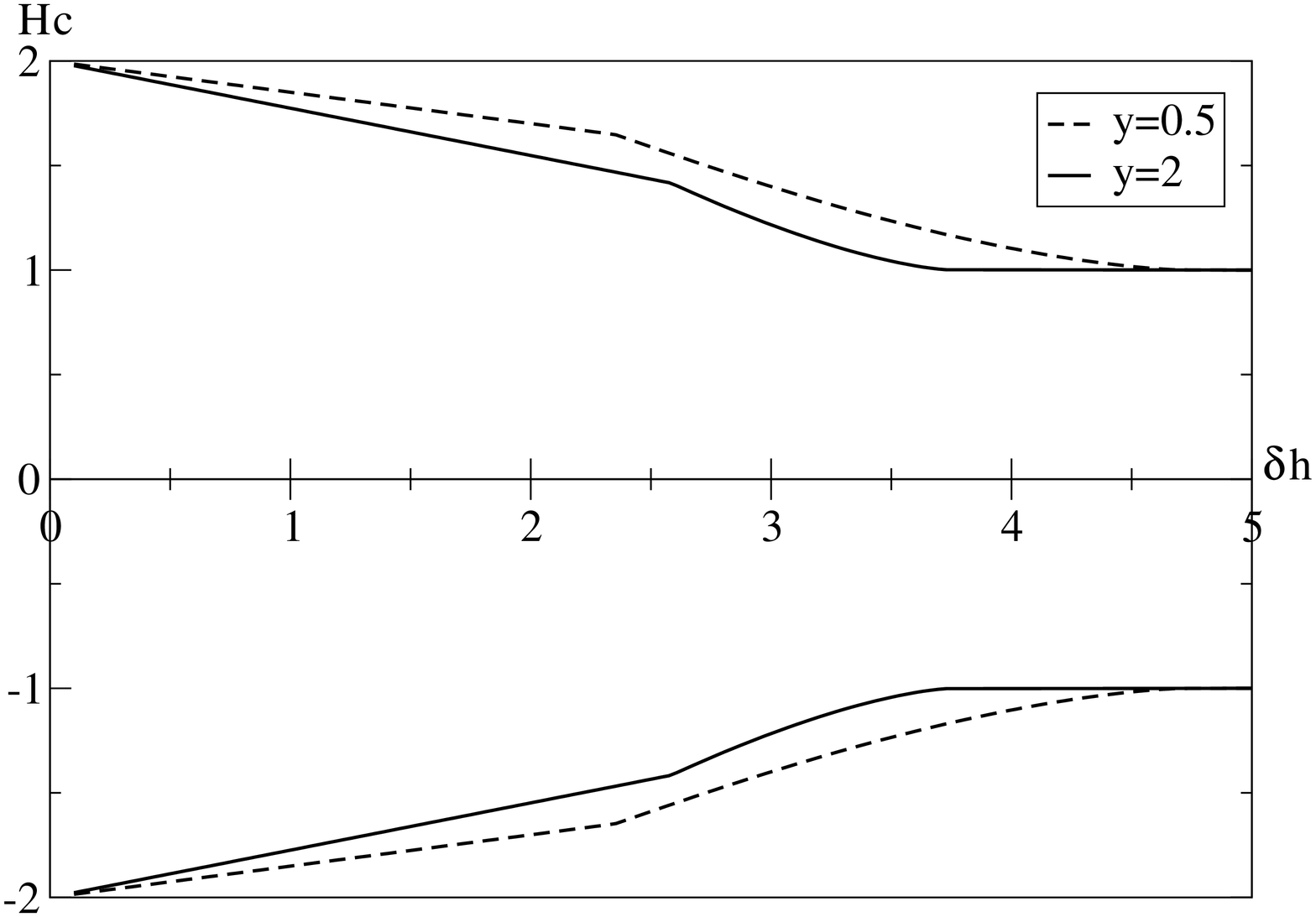}}} \par}
{\centering \resizebox*{0.68\columnwidth}{!}{
\rotatebox{270}{\includegraphics{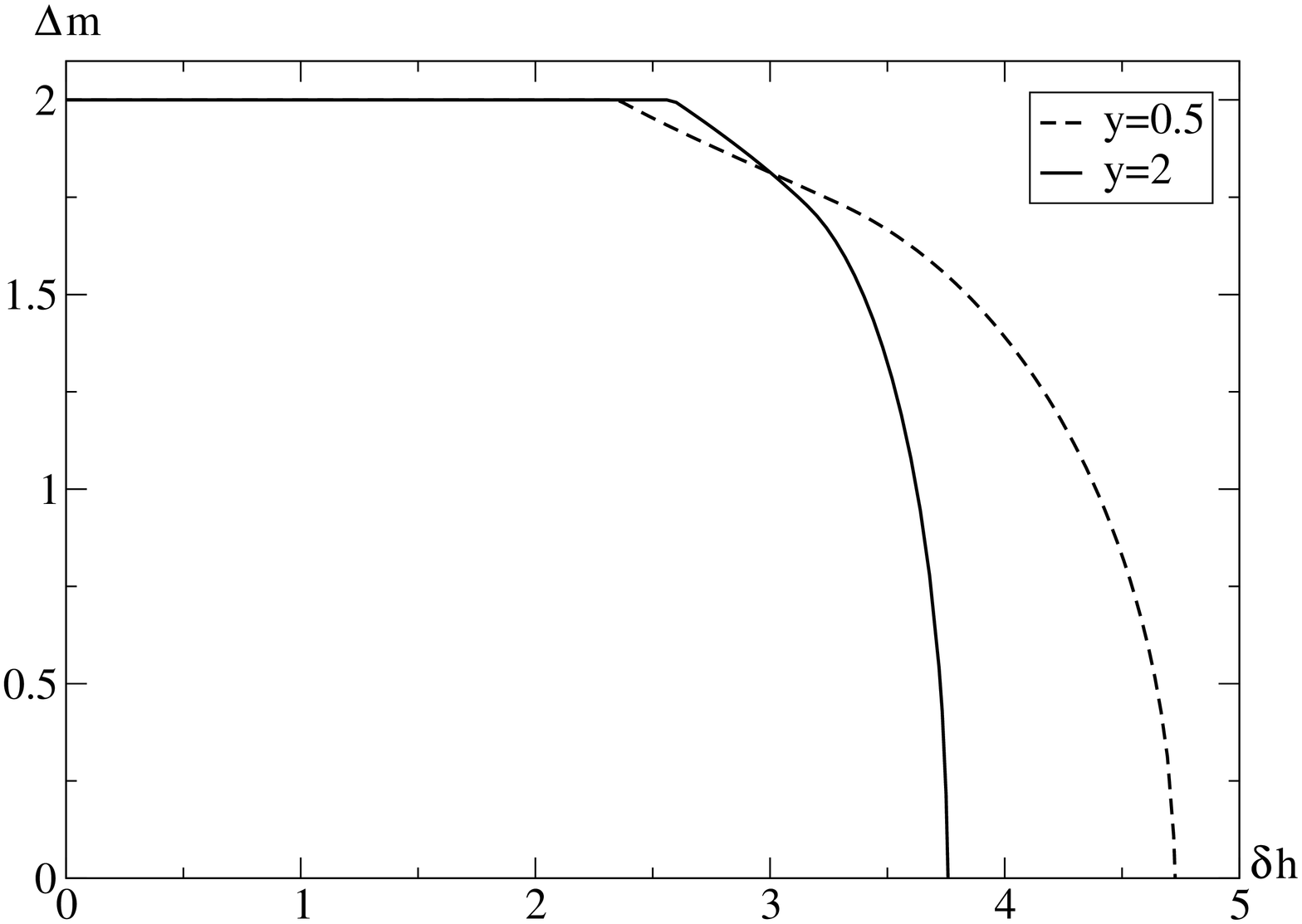}}} \par}

\caption{\textbf{Top} The phase diagram for the non-equilibrium RFIM on a
Cayley tree with coordination number \protect\( \alpha =3\protect \)
using the distribution (\ref{eq:poly_dist_1}) and taking the exchange
constant \protect\( J=1\protect \). The dotted line is for \protect\( y=0.5\protect \)
and solid line is for \protect\( y=2\protect \). The initial, linear
part, of the phase boundary is due to the finite cutoff of the distribution
(\ref{eq:poly_dist_1}). There is a discontinuity in slope of \protect\( H_{c}(\delta h)\protect \)
at the equilibrium critical disorder \protect\( \delta h_{c}\protect \).
\textbf{Bottom} The magnetization jump for the RFIM on Cayley trees
for \protect\( z=4\protect \) and the distribution (\ref{eq:poly_dist_1}),
with the exchange constant \protect\( J=1\protect \). The dotted
line is for \protect\( y=0.5\protect \) while the solid line is for
\protect\( y=2\protect \). In both cases we find the same critical
exponent, for example \protect\( \beta =1/2\protect \). In contrast,
the mean-field result is \protect\( \beta =1/y\protect \).}

\label{fig2}
\end{figure}
The coordination number of a tree is taken to be \( z \), while the
probability that a spin is up is \( P_{+} \) and the probability
that a spin is down is \( P_{-} \). The probability that a spin is
up at level \( l \) can be written in terms of the probabilities
at the level which is one lower down in the tree, this yields \cite{bruinsma84a,dhar97a}\begin{equation}
\label{eq:recursion_cayley_tree}
P_{+}(l)=\sum ^{\alpha }_{g=0}\left( \begin{array}{c}
\alpha \\
g
\end{array}\right) P^{g}_{+}(l-1)P^{\alpha -g}_{-}(l-1)a_{+}(\alpha ,g)
\end{equation}
 where \( a_{+}(\alpha ,g) \) is the probability that the local effective
field is positive when \( g \) neighbors are up. If we know the distribution
\( \rho (h_{i}) \) we can compute \( a_{+}(\alpha ,g) \). Analyzing
the equilibrium behavior, we have,

\begin{equation}
\label{eq:factors}
a^{eq}_{+}(\alpha ,g)=\int ^{\infty }_{(\alpha -2g)J-H}\rho (h)dh
\end{equation}

The equilibrium Cayley tree model has been extended to the non-equilibrium
case by considering a growth problem in which the spin \textit{above}
the currently considered level in the tree is pinned in the down position
\cite{dhar97a,sabhapandit00a}. This models the growth of a domain.
The formalism is the same as in Eq.~(\ref{eq:recursion_cayley_tree}),
with the modification that \begin{equation}
a_{+}^{neq}(\alpha ,g)=a_{+}^{eq}(z,g).
\end{equation}
 From this equality and the form (\ref{eq:recursion_cayley_tree})
it is easy to derive all of the non-equilibrium results from the equilibrium
results found using Eqs.~(\ref{eq:recursion_cayley_tree}) and (\ref{eq:factors}).
To find the hysteresis curve on a Cayley tree, we just shift the equilibrium
magnetization as a function of field: by \( H\rightarrow H-J \) when
sweeping from large positive fields and; by \( H\rightarrow H+J \)
when sweeping from large negative fields. The behavior is evident
in previous numerical work, but does not seem to have been noticed
before.

By direct iteration of the recurrence relation (\ref{eq:recursion_cayley_tree})
we show that a stable steady state solution, \( P^{*}_{+}=1-P^{*}_{-} \),
exists. It is easy to solve equation (\ref{eq:recursion_cayley_tree})
in the steady state limit, at least for small values of \( \alpha  \).
For \( \alpha =1,\, \, 2 \) Cayley trees have no ordered state for
any finite \( \delta h \), for the disorder distribution (\ref{eq:poly_dist_1}).
But for \( \alpha =3 \) a ferromagnetic state does exist for a range
of disorder. As we see from Eq.~(\ref{eq:recursion_cayley_tree}),
the \( \alpha =3 \) case leads to a polynomial of order \( 3 \)
which can be simplified to, \begin{equation}
\label{eq:steady_state_solution}
\frac{m}{4}[m^{2}(1-3b+a)-1+3a+3b]=0
\end{equation}
were \( m=2(P_{+}^{*}-1/2) \), \( a=a^{eq}_{+}(3,0) \) and \( b=a^{eq}_{+}(3,1) \).
Eq.~(\ref{eq:steady_state_solution}) has the following solutions:
\begin{equation}
\label{eq:general_solution_of_steady_state_magnetization}
m=0;\qquad {\textrm{and}}\, \, \, \, m=\pm \left( \frac{3a+3b-1}{3b-1-a}\right) ^{1/2}.
\end{equation}
 These solutions apply for any disorder distribution. For the distribution
\( \rho _{1}(h) \), performing the integrals yields, \begin{equation}
\label{eq:magnetization_for_rho1}
m=\left[ \frac{4y-12(y+1)\overline{J}+3(3^{y+1}+1)\overline{J}^{y+1}}{3(1-3^{y})\overline{J}^{y+1}}\right] ^{1/2}
\end{equation}
 We can now expand the magnetization around the critical point, \( J_{c} \),
\( \overline{J}=\overline{J}_{c}-\epsilon  \). We find, \begin{equation}
m\sim \left[ (-12(y+1)+3\left( \begin{array}{c}
y+1\\
y
\end{array}\right) (1+3^{y+1})\overline{J}^{y})\epsilon \right] ^{1/2}
\end{equation}
 Thus \( m\sim \epsilon ^{1/2} \) for any \( y \), so that \( \beta =1/2 \)
is universal. Since the non-equilibrium behavior on trees is related
to that of the equilibrium behavior in such a simple manner, this
universality extends to the hysteresis and avalanche exponents. It
is easy to confirm numerically that the behavior extends to large
values of the branch co-ordination number \( \alpha  \). Moreover
by doing an expansion of (\ref{eq:recursion_cayley_tree}) using \( P_{+}=1/2+m \),
it is possible to show analytically that only the first and third
order terms in \( m \) exist, regardless of the value of \( y \)
in the disorder distribution (\ref{eq:poly_dist_1}). This confirms
that for this distribution, the behavior is universal for all coordination
numbers.

For the distribution \( \rho _{2}(h) \) and \( \alpha =3 \) we get
from Eq.~(\ref{eq:general_solution_of_steady_state_magnetization})\begin{equation}
\label{eq:magnetization_for_rho2}
m=\left[ \frac{4-3(3^{y+1}+1)\overline{J}^{y+1}}{3(3^{y}-1)\overline{J}^{y+1}}\right] ^{1/2}.
\end{equation}
Just like we have done before we can expand \( m \) around the critical
point, \( \overline{J}=\overline{J_{c}}-\epsilon  \):\[
m\sim \left[ \left( \begin{array}{c}
y+1\\
y
\end{array}\right) (3^{y+1}+1))\overline{J_{c}}^{y}\epsilon \right] ^{1/2}.\]
Thus for \( \rho _{2} \) \( \beta =1/2 \) is a universal exponent,
too. 

In summary, on complete graphs (i.e. in mean-field theory) 
the RFIM at $T=0$ is non-universal.
In particular, the stretched exponential disorder distribution leads
to a non-universal order parameter exponent and 
non-universal integrated avalanche exponent.  In addition,
the power law distribution has a regime in which 
a predominance of small random fields destroys the
transition and the RFIM always has a finite 
magnetization.
In contrast the Cayley tree does not show
either of these behaviors.  
Even in the limit of large coordination it is 
universal, with the usual mean-field order-parameter
exponent $1/2$.  We have carried out some 
preliminary numerical studies of the behavior in
three dimensions (with short range
interactions) and find that the power law
distribution of random fields does {\it not}
destroy the transition. Moreover, 
the exceedingly small value of $\beta$ in three
dimensions renders any non-universality in $\beta$ a moot
point. However the behavior in dimensions higher than
three, or for longer range interactions in 
three dimensions could be more interesting.
Finally,  even for short range interactions in
three dimensions, there have been suggestions of non-universality
in the finite size scaling behavior \cite{angles97a}. 
It is unclear, as yet, whether that behavior is related
to the non-universality seen here.

\begin{acknowledgments}
This work has been supported by the DOE under contract DE-FG02-90ER45418.
\end{acknowledgments}

\end{document}